\documentclass[]{interact}
\usepackage{epstopdf}
\usepackage[caption=false]{subfig}
\usepackage[natbibapa,nodoi]{apacite}
\theoremstyle{plain}

\usepackage[colorlinks=true, citecolor=blue, linkcolor=blue, urlcolor=blue]{hyperref}

\newcommand{\mt}{\mathit{MT}}
\newcommand{\er}{\mathit{ER}}
\newcommand{\id}{\mathit{ID}}
\newcommand{\aic}{\mathit{AIC}}
\newcommand{\mae}{\mathit{MAE}}

\begin{document}

\articletype{Research Article. Word count: 7{,}094}

\title{Blur Effects on User Performance in Target-Pointing Tasks}

\author{
\name{Ryuto Tomihari\textsuperscript{a}, Taiki Kinoshita\textsuperscript{a}, Yosuke Oba\textsuperscript{a}, Shota Yamanaka\textsuperscript{b,$\dagger{}$}, Homei Miyashita\textsuperscript{a}}
\affil{\textsuperscript{a} Meiji University, Nakano-ku, Tokyo, Japan; \textsuperscript{b}\thanks{${}^\mathrm{\dagger{}}$\ Corresponding author: S. Yamanaka. LY Corporation Research, LY Corporation, Kioi Tower, 102-8282, Tokyo Garden Terrace Kioicho, 1-3, Kioi-cho, Chiyoda-ku, Tokyo, Japan. Email: syamanak@lycorp.co.jp} LY Corporation, Chiyoda-ku, Tokyo, Japan}
}

\maketitle

\begin{abstract}
In projectors and head-mounted displays, an out-of-focus image appears blurred.
Even when a display itself is in focus, computer operation may be hindered if the display is far from the user or if a user has poor visual acuity, because the user cannot see the screen clearly.
In this study, we conducted an experiment in which participants performed a pointing task under blurred display conditions and investigated the relationship between blur strength and user performance.
The results showed that movement time and error rate increased as blur became stronger, and that the effect of blur on movement time was larger when targets were smaller.
We further showed that movement time can be estimated with high accuracy by a model that improves on Fitts' law.
In a follow-up experiment to examine the applicability of this model, we adjusted target size for each participant and showed that the effect of blur level on movement time could be reduced.
These findings suggest potential use in tools that adapt user interfaces to users' visual acuity.
\end{abstract}


\section{Introduction}
\label{sec:intro}
Pointing at targets such as icons and hyperlinks is one of the most fundamental actions in user interface (UI) operation on PCs, smartphones, virtual reality (VR) systems, and other computing platforms.
In the field of HCI, ISO 9241-411, a multidirectional tapping task based on Fitts' law \citep{Fitts54}, is often adopted in user studies to measure pointing performance \citep{iso2012}.

In experiments using PCs or touchscreens, participants are generally expected to operate the system from an appropriate distance from a display that is properly rendered, and researchers pay attention to creating a setup in which participants can comfortably view the screen.
This assumption is implicitly reflected in the common statement in research papers that ``all participants had normal or corrected-to-normal vision'' \citep{Kim20,Jiang14,Minakata19,Bieg09,Wilson05,Yamanaka18mobilehci}.
This suggests that researchers recognize the importance of participants being able to clearly see both the targets and the mouse cursor.

However, in realistic environments, the screen may not always be clearly displayed, and even when it is, users may not be able to perceive it accurately.
For example, pointing performance has been reported to decrease when the user is far from the display (an environmental factor) \citep{Wang13,Hourcade12}, or when the images presented to the left and right eyes in a head-mounted display (HMD) are not perfectly in focus (a device factor) \citep{Batmaz22,Barrera19}.
Reduced visual acuity can also affect performance (a user factor) \citep{Jacko00a,MyopiaPointing}.
As diverse users and a wide variety of display devices, such as mobile devices, projectors, and HMDs, have become more common, researchers have examined situations in which the screen cannot always be clearly visible to the user.

However, previous studies have not clarified the quantitative relationship between reduced display clarity and UI operation performance.
For example, if the degree of blur doubles according to some metric, does operational time also double, or is the relationship different?
To fill this gap in understanding, this study quantitatively evaluates the effect of reducing display clarity on pointing performance by conducting ISO 9241-411 tasks while Gaussian blur is applied to the screen.
Please refer to the following URL for a system demo video with blur applied to the target and cursor (\url{https://drive.google.com/file/d/1CSUKq0Cy7yiqSdUlfSSw5B3NEng_4l6a/view?usp=sharing}).
Furthermore, because conventional performance models based on Fitts' law could not estimate movement time accurately, we also aim to improve the model.
As an application of the revised model, we propose a method for reducing blur-induced increases in movement time by enlarging target size, and we conducted a follow-up experiment to test its effectiveness.

Our contributions are threefold.
First, we conducted an experiment in which display clarity was quantitatively reduced through Gaussian blur strength and showed that this degrades pointing performance.
Second, we derived a mathematical model that estimates movement time with high accuracy regardless of blur strength.
Third, through a follow-up experiment, we evaluated the usefulness of the model for adjusting target size so as to reduce movement-time increases.

\section{Related Work}
\subsection{Pointing Performance and Fitts' Law}
The movement time $\mt$ required to select a target on a display with a mouse cursor can be estimated by Fitts' law \citep{Fitts54,MacKenzie92}.
\begin{equation}
    \mbox{One-Part Model: }\mt = a + b \log_2{\left(\frac{A}{W}+1\right)}
    \label{eq:fitts}
\end{equation}
Here, $A$ is the distance to the target center, and $W$ is the target size.
Throughout this paper, italic lowercase letters $a$--$e$ denote empirically determined constants.
The logarithmic term is called the index of difficulty $\id$.

According to Fitts' law, $\mt$ is predicted to remain unchanged when $A$ and $W$ vary in the same proportion.
In some situations, however, the effects of $A$ and $W$ on $\mt$ differ in magnitude.
For example, long-distance movement may be easy while fine cursor control remains difficult, in which case the following Two-Part Model provides a better fit \citep{Graham95,welford1968a,Janzen16,Shoemaker12}.
\begin{equation}
    \mbox{Two-Part Model: }\mt =  a + b \log_2{(A)} - c \log_2{(W)}
    \label{eq:2part}
\end{equation}
We will also compare this model because blur is expected to make the final operation of moving the cursor into the target area of size $W$ more difficult than simply moving the cursor over distance $A$.

\subsection{Computer Operation Under Reduced Visual Clarity}
\subsubsection{Device Factors}
A typical device factor is the blurred display caused by an improperly focused projector.
For example, the manual for a BenQ projector states that the image may become blurred when the distance between the lens and the screen exceeds 3.2~m \citep{benq_faq_blurry_image_kn00070}.
The manual for an Epson projector also instructs users to project a test pattern and adjust the focus ring when the projected image is blurry \citep{epson_en_blurry_troubleshoot}, indicating that blur is a common issue in projector use.

Regarding visual presentation in VR environments, many HMDs have a fixed focal distance, which causes a vergence--accommodation conflict (VAC) because the vergence and accommodation cues of displayed objects may not match, thereby degrading pointing performance \citep{Batmaz22}.
In stereoscopic displays, depth perception can be compressed and the pointing accuracy in depth is reduced \citep{Barrera19}.
Visual conflicts may also arise because the focal plane of the user's real finger and that of the virtual object do not match, which causes one of them to appear blurred \citep{Bruder13}.

\subsubsection{Environmental Factors}
Visual clarity can change depending on environmental conditions such as viewing distance, ambient lighting, and reflections, and pointing performance may therefore change.
For example, when the distance between the user and the display changes, performance and subjective comfort can deteriorate even if the apparent target size is kept constant, such as by enlarging targets on the screen in proportion to the increase in distance \citep{Wang13,Hourcade12}.

Lighting and glare are also factors that impair visibility, and many computer users have probably experienced difficulty seeing a PC or smartphone screen because sunlight or room lights were reflected on it.
In an experiment by \cite{Hamedani20}, participants performed a document-reading task on a screen with glare, and gaze analysis revealed negative effects such as higher fixation rates and lower blink rates.
When a room is too bright relative to the brightness of a projector, the screen also becomes difficult to see, which increases response time and reduces accuracy in a visual search task \citep{Gu24}.
Although we found no prior work focusing specifically on pointing performance, these studies show that environmental conditions can alter visual quality and hinder computer-based work.

\subsubsection{User Factors}
Low vision and ophthalmic conditions can also change the quality of visual feedback and thereby affect UI operation performance.
\cite{MyopiaPointing} conducted a touch-pointing task and showed that participants with amblyopia required significantly longer time than participants with myopia or normal vision.
\cite{Jacko00a} and \cite{Jacko00b} examined cursor-movement efficiency in low-vision users and showed that reduced visual acuity lengthens $\mt$ and lowers performance compared with that of users with better vision.
These findings are also supported by research in ophthalmology, which has shown that visual acuity and contrast sensitivity affect both time and accuracy in icon-selection tasks \citep{Scott02a}.

User studies that simulate reduced visual acuity can be conducted either by having participants wear mismatched prescription glasses \citep{lowVisionGoggle} or by applying blur through image processing \citep{bokesimuEnglish,Gilbertson2015}.
Because the use of eyeglasses may cause symptoms such as headaches or dizziness, we use image processing, namely Gaussian blur, in our experiment.
Gaussian blur has been reported to reasonably reproduce both projector defocus \citep{Nagase11,Brown06,Galetto22,Oyamada08} and reduced visual acuity, including its optical frequency characteristics \citep{bokesimuEnglish,Mori11LowVision}.

\section{Experiment 1: Examining the Effect of Blur on Pointing Performance}
\subsection{Research Ethics}
This study involved only PC operation tasks and did not collect sensitive personal data.
In accordance with the ethics guidelines of Meiji University, we confirmed that such procedures do not require formal review by the institutional review board.
Participants provided verbal informed consent and were informed of their right to withdraw at any time without penalty.
Data were stored and analyzed in anonymized form.
The same ethical protocols and data handling procedures were followed in Experiment 2.

\subsection{Participants}
Twelve students aged 20--26 years participated in the experiment (mean age $=$ 21.6 years, SD $=$ 2.11).
All participants reported that they normally used a mouse with their right hand, and thus all participants operated the mouse with their right hand in this experiment.
They had normal or corrected-to-normal vision.
Three wore glasses, three wore contact lenses, and the remaining six used no correction.

\subsection{Apparatus}
We used a desktop PC (Core i9-12900KF, GeForce RTX 3070Ti, 48.0 GB RAM, Windows 10 Pro), a 2560$\times$1440-pixel display (PHILIPS 328P6A, 31.5 inches), and a Logitech mouse (G300s, 1000 dpi).
The cursor size and speed were set to the Windows default.
The experimental system was implemented in Hot Soup Processor and presented in full-screen mode.
As in projector defocus and low-vision conditions, Gaussian blur was applied to the entire screen.

\subsection{Task}
\begin{figure*}[t]
    \centering
    \includegraphics[width=1\linewidth]{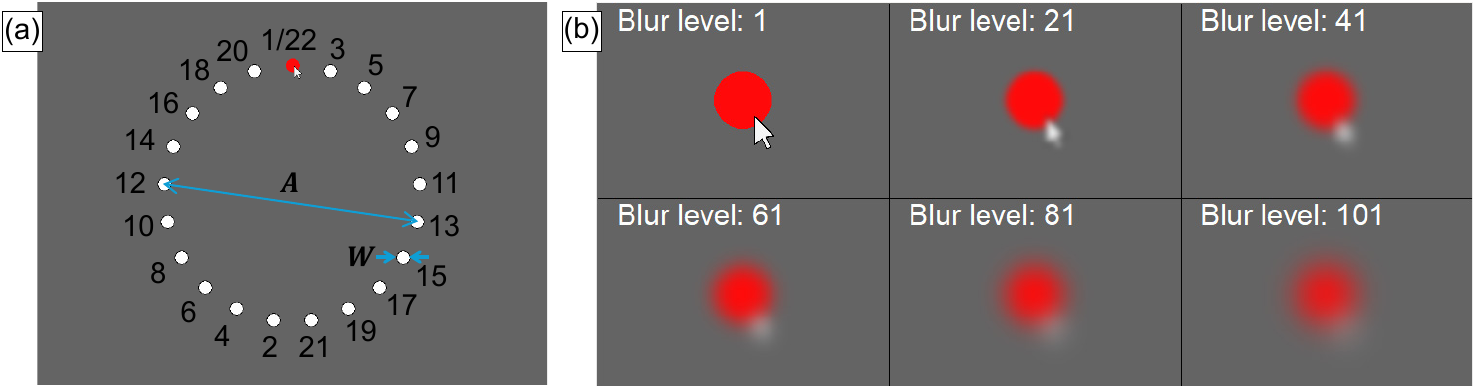}
    \caption{(a) Screen used in Experiment 1, in which the targets were clicked in numerical order, and (b) how the cursor appeared near the target under the six blur levels.}
    \label{fig:studyScreen}
\end{figure*}

Following ISO 9241-411 \citep{iso2012,Soukoreff04}, the task was to click 21 circles in the order shown in Figure~\ref{fig:studyScreen}a.
The topmost circle was the start target, and when the participant clicked the red target, the next target to be selected became red.
For a fixed $A\times W$ condition, the 21 trials following the selection of the start target constituted one \textit{session}.

A success sound was played when the target was clicked, and a failure sound was played when the target was missed.
If the click failed, the participant had to continue aiming at the same target until selection succeeded.
Although some previous studies have added visual feedback such as target blinking after a failure \citep{Yamanaka21hcomp}, we avoided visual feedback in our experiment so as not to reveal the position of a blurred target.

\subsection{Gaussian Blur}
Gaussian blur applies blur to an input image, which in our experiment was the entire screen, by convolving it with a kernel based on a Gaussian function and suppressing high-frequency components.
We used OpenCV's \texttt{GaussianBlur} function, for which the blur strength $\sigma$ and the kernel size $\mathit{ksize}$, which defines the computation range for the blur and must be an odd integer, can be specified.
Because specifying one of these automatically determines the other appropriately, we specified only $\mathit{ksize}$ in our system and treated it as the blur strength.
In this case, $\sigma = 0.3\times[(\mathit{ksize}-1)/2 - 1] + 0.8$.
For replicability, readers are directed to the official documentation for the detailed computation method \citep{opencvGaussianBlur}.

\subsection{Design}
This study used a $2\times 4\times 6$ within-participants design.
We used two $A$s (300 and 500 pixels), four $W$s (12, 18, 36, and 78 pixels), and six levels for blur $B$ (1, 21, 41, 61, 81, and 101) specified as $\mathit{ksize}$ in pixels; see Figure~\ref{fig:studyScreen}b.
The $\id$ ranged from 2.28 to 5.42 bits.
When we attempted to further increase $\id$ by reducing $W$, the task sometimes became extremely difficult to continue under strong blur, so we set 5.42 bits as the upper limit.

\subsection{Procedure}
For each $B$ condition, participants first performed one practice session and then eight data-collection sessions ($2_A\times4_W$).
In the practice session, we used a medium difficulty condition that did not appear in the data-collection sessions ($A=400$, $W=23$ pixels, $\id=4.20$ bits).
Because each participant experienced every combination of conditions, the data collection consisted of $2_A\times4_W\times6_B\times21_\mathrm{repetitions}=1{,}008$ trials.
The order of $B$ conditions was randomized across participants using a Latin square, and the order of the $2_A\times4_W=8$ conditions was also randomized.

Participants were allowed to take breaks between sessions.
We instructed them to perform as quickly and accurately as possible \citep{MacKenzie92,Soukoreff04} and not to lift the mouse during a session (i.e., no clutching) \citep{Nancel15clutch,Casiez07rubber}.
The experiment took approximately 60 minutes per participant.

After the experiment, we conducted a NASA Task Load Index (NASA-TLX) questionnaire \citep{NASA} and a free-description questionnaire to evaluate subjective workload and perceived difficulty.
For NASA-TLX scoring, either the weighted workload or the simple mean (raw-TLX) is commonly used \citep{RTLX}.
Because weighted workload requires pairwise comparison of all six items and thus imposes a large burden on participants, we used raw-TLX, especially given its high correlation with weighted workload ($0.93 < r < 0.98$ \citep{RTLXiiyo,Miyake93,Miyake95}).

For subjective difficulty, participants rated each $B$ condition with an integer score from 0 (easy, can be operated without any problem) to 100 (difficult, very hard to operate).
They provided ratings while viewing a window that displayed all $B$ conditions side by side and allowed them to move the mouse cursors simultaneously (Figure~\ref{fig:studyScreen}b).

\section{Results of Experiment 1}

\begin{table}[t]
    \centering
    \caption{Results of the RM-ANOVA for $\er$ and $\mt$ in Experiment 1.}
    \label{table:e1ERMT}
    \resizebox{1.0\textwidth}{!}{
    \begin{tabular}{l|lll|lll}
        &\multicolumn{3}{c|}{$\er$}&\multicolumn{3}{c}{$\mt$}\\
        \multicolumn{1}{c|}{Factors} & \multicolumn{1}{c}{$F$} & \multicolumn{1}{c}{$p$} & \multicolumn{1}{c|}{$\eta_{p}^{2}$} & \multicolumn{1}{c}{$F$} & \multicolumn{1}{c}{$p$} & \multicolumn{1}{c}{$\eta_{p}^{2}$} \\
        \hline
            $A$ & $F(1,11)=0.005086$ &0.944 & 0.000& $F(1,11)=80.62$ & $<0.001$ & 0.880 \\
            $W$ & $F(3,33)=226.0$ & $<0.001$ &0.954 & $F(3,33)=91.39$ & $<0.001$&0.893 \\
            $B$ & $F(5,55)=165.4$ & $<0.001$& 0.938& $F(5,55)=13.55$ &$<0.001$& 0.552\\
            \hline
            $A\times W$ & $F(3,33)=0.5897$ & 0.626& 0.051& $F(3,33)=0.6050$ & 0.616& 0.052\\
            $A\times B$ & $F(5,55)=1.291$ & 0.281& 0.105& $F(5,55)=0.2494$ & 0.938& 0.022\\
            $W\times B$ & $F(15,165)=36.50$ & $<0.001$& 0.768& $F(15,165)=2.034$ &$<0.05$ & 0.156\\
            \hline
            $A\times W\times B$ & $F(15,165)=0.5755$ & 0.891& 0.050& $F(15,165)=0.3572$ & 0.987&0.031 \\
            \hline
    \end{tabular}
    }
\end{table}

We recorded data from 12{,}096 trials obtained from 12 participants.
In prior studies on Fitts' law, outliers have been removed based on click coordinates or $\mt$ \citep{Findlater17,MacKenzie08,Soukoreff04}.
However, in our experiment, especially under strongly blurred conditions, it is not surprising that participants occasionally clicked far from the target, or produced large $\mt$ values because they pointed very cautiously.
Rather, our interest lies precisely in analyzing the error rate $\er$ and $\mt$ including such behavior, so we decided to include all trials in the analysis.
Because many prior studies have also conducted analyses without outlier removal \citep{Fitts54,MacKenzie92,Accot03}, this decision does not provide clear grounds for considering our analysis inappropriate.

The independent variables were $A$, $W$, and $B$, and the dependent variables were $\er$ and $\mt$.
$\er$ was defined as the proportion of trials in which the first click did not succeed, and $\mt$ was defined as the elapsed time from clicking the previous target to the first click on the current target.
In this paper, error bars in graphs represent standard errors, and *, **, and *** indicate $p<0.05$, $p<0.01$, and $p<0.001$, respectively.
Because ANOVA is robust to violations of the normality assumption \citep{Blanca17,Schmider10,Tsandilas24}, we consistently used RM-ANOVAs.
Bonferroni correction was applied to adjust $p$-values in pairwise tests.

\subsection{Error Rate}
\begin{figure}[th]
    \centering
    \includegraphics[width=0.8\linewidth]{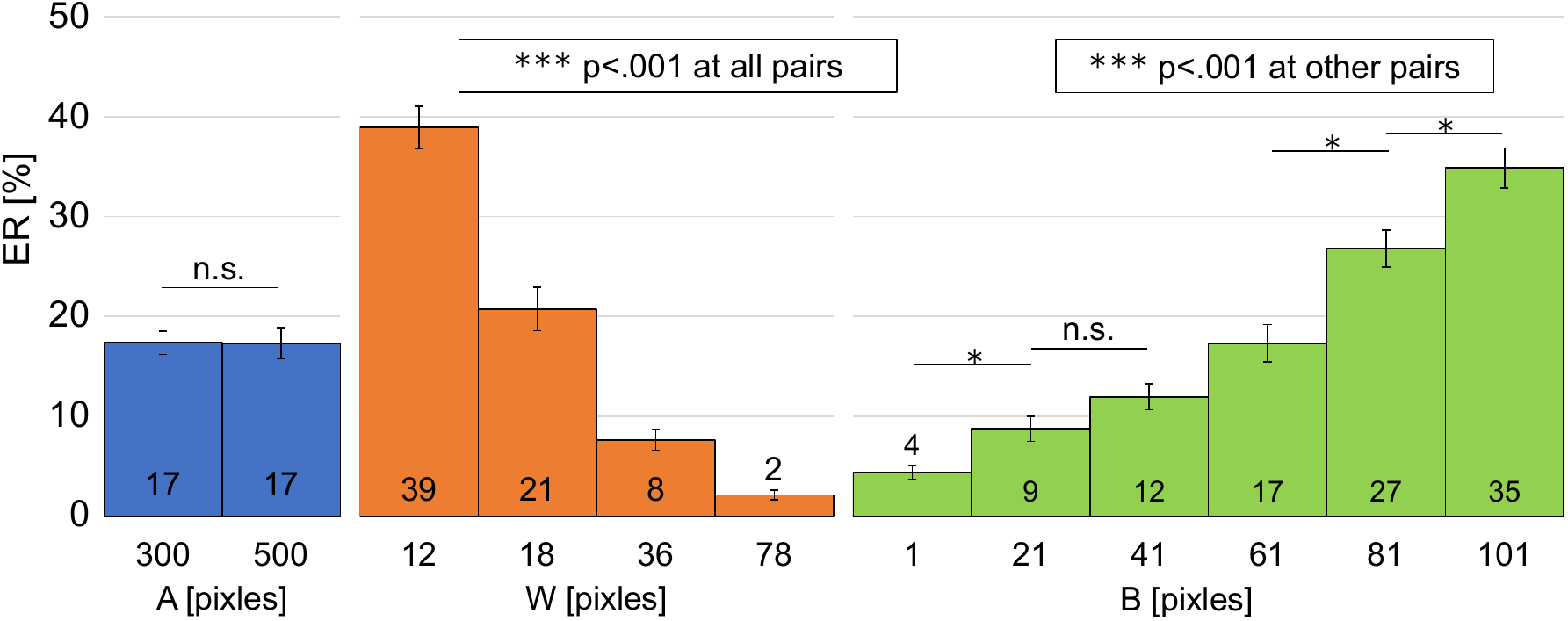}
    \caption{Effects of $A$, $W$, and $B$ on $\er$ in Experiment 1.}
    \label{fig:e1ER}
\end{figure}

Errors occurred in 2{,}383 trials out of 12{,}096, yielding an overall $\er$ of 19.7\%.
Table~\ref{table:e1ERMT} shows the ANOVA results.
Significant main effects were found for $W$ ($p<0.001$) and $B$ ($p<0.001$), but not for $A$ ($p=0.944$) (Figure~\ref{fig:e1ER}).
$\er$ increased as $W$ became smaller and as $B$ became larger.

\begin{figure}[th]
    \centering
    \includegraphics[width=\linewidth]{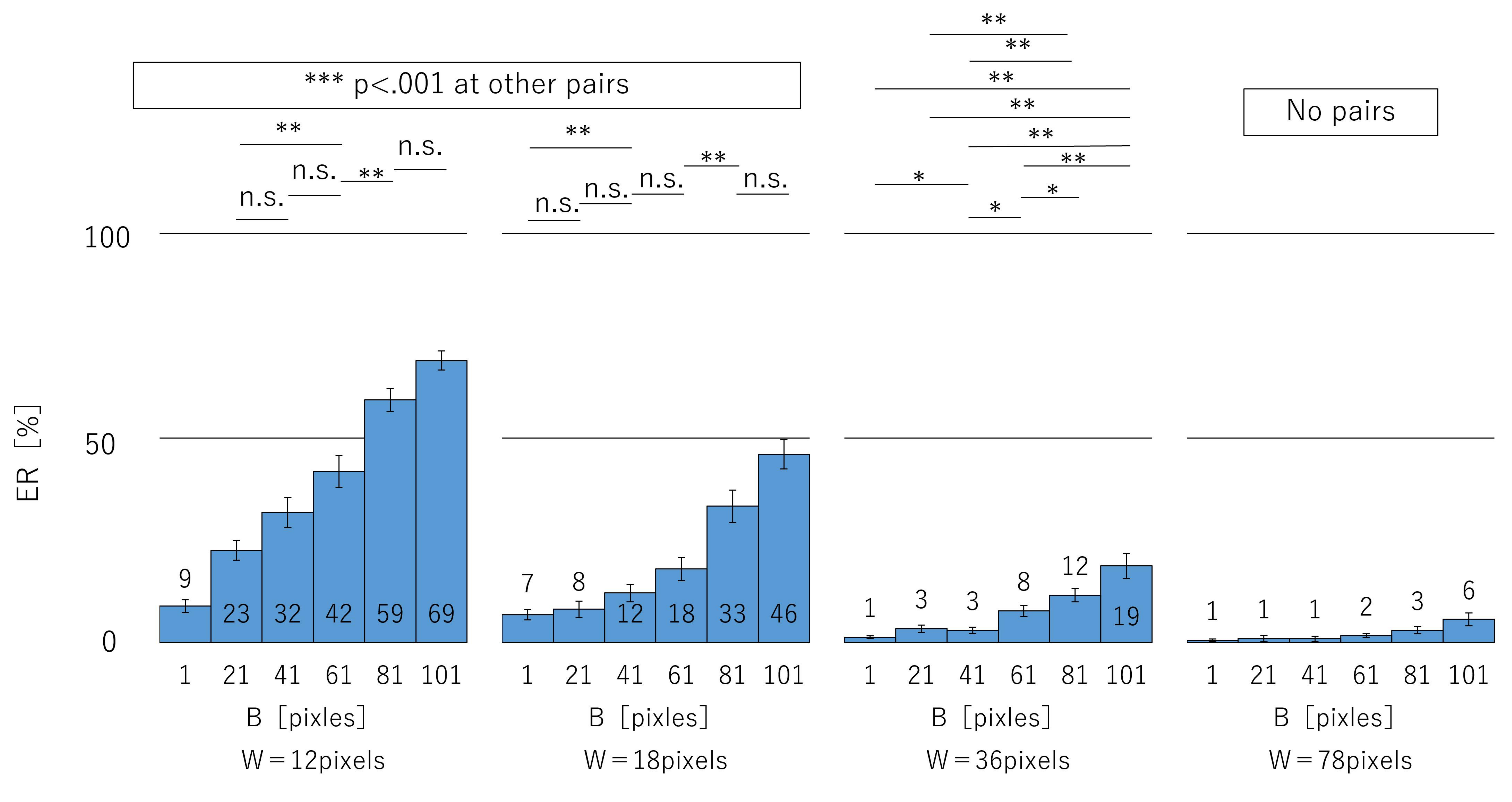}
    \caption{Effect of $B$ on $\er$ for each $W$ in Experiment 1.}
    \label{fig:e1ER_wsp}
\end{figure}

A significant interaction was found for $W \times B$ ($p<0.001$).
No interaction was found for $A \times W$ ($p=0.626$) or $A \times B$ ($p=0.281$).
Figure~\ref{fig:e1ER_wsp} shows the effect of $B$ for each $W$.
When the target was large ($W=78$ pixels), $\er$ was at most 6\%, whereas the effect of blur became stronger as the target became smaller.


\begin{figure}[th]
    \centering
    \includegraphics[width=0.9\linewidth]{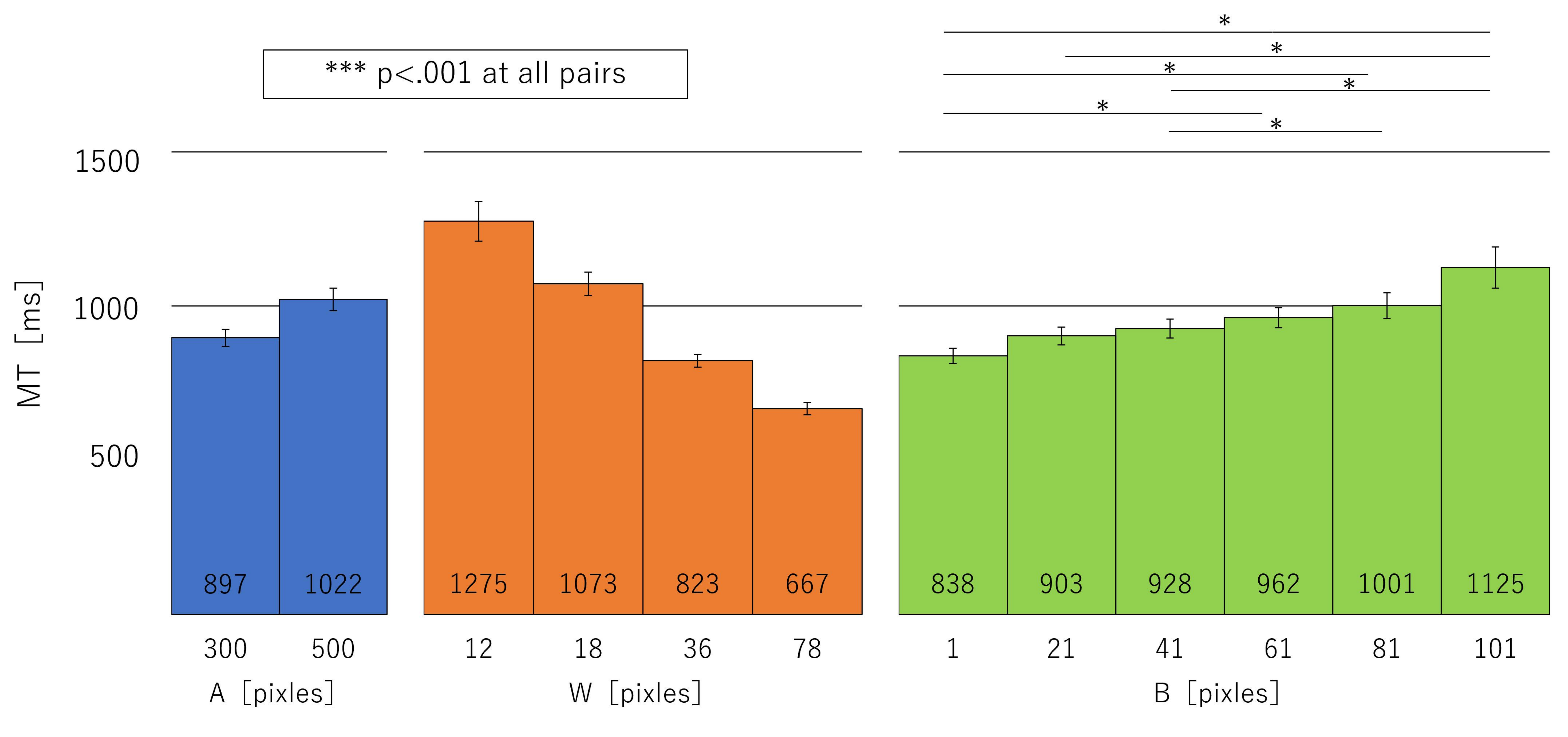}
    \caption{Effects of $A$, $W$, and $B$ on $\mt$ in Experiment 1.}
    \label{fig:e1MT}
\end{figure}
\begin{figure}[th]
    \centering
    \includegraphics[width=\linewidth]{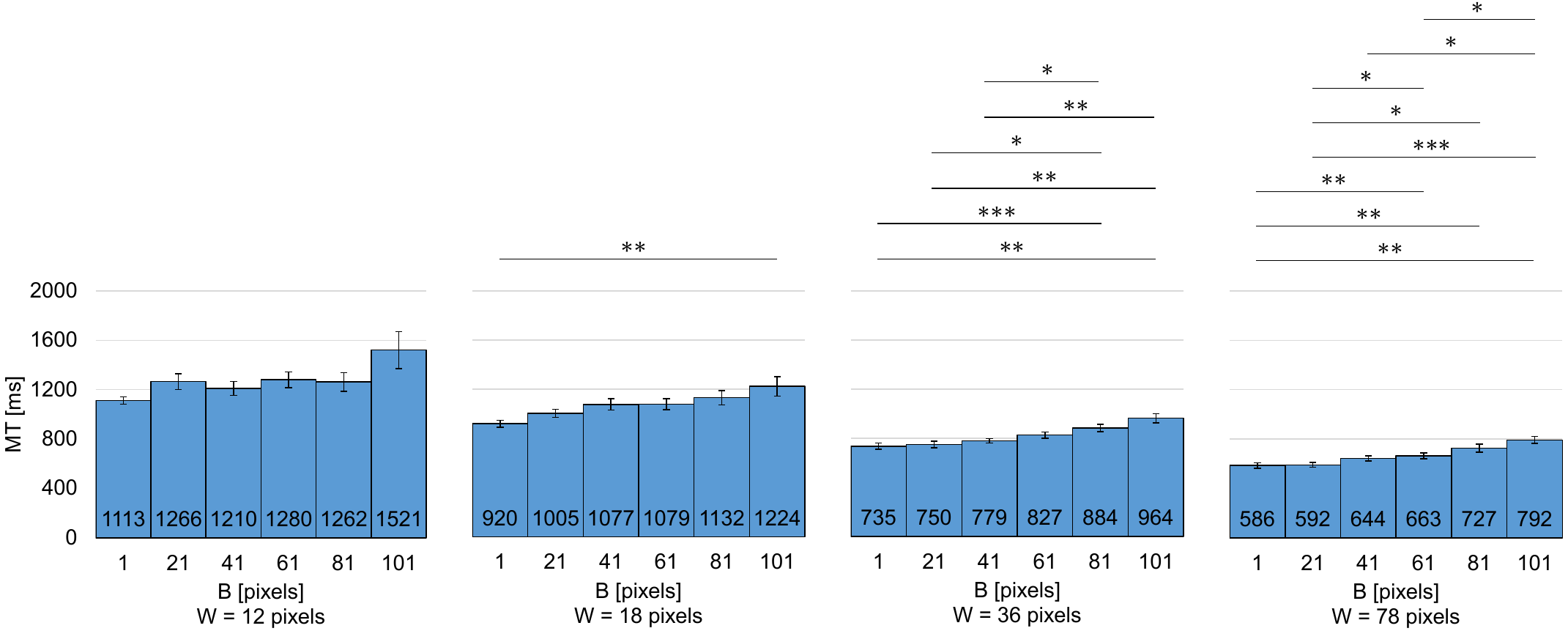}
    \caption{Effect of $B$ on $\mt$ for each $W$ in Experiment 1.}
    \label{fig:e1MT_WandB}
\end{figure}

\subsection{Movement Time}

The analysis of $\mt$ used data from 9{,}713 error-free trials \citep{Accot03,Bi13a,Ko20FF2D,Yamanaka18gi}.
The overall mean $\mt$ was 959 ms.
Significant main effects were found for $A$ ($p<0.001$), $W$ ($p<0.001$), and $B$ ($p<0.001$) (Figure~\ref{fig:e1MT}).

A significant interaction was found for $W \times B$ ($p<0.05$).
As shown in Figure~\ref{fig:e1MT_WandB}, when $W=12$ pixels, error bars were large and no significant differences were found between $B$ conditions, but significant differences between $B$ conditions became more common as $W$ increased.
Importantly, the difference in $\mt$ between $B$ conditions was larger when $W$ was smaller.
Specifically, when $W=78$ pixels, $\mt$ increased from 586 to 792 ms depending on the $B$ condition, which is an increase of 206 ms.
In contrast, when $W=12$ pixels, $\mt$ increased from 1{,}113 to 1{,}521 ms; an increase of 408 ms.
Thus, the change in $\mt$ was 98\% larger ($100\%\times(408-206)/206$).
This indicates that the effect of $B$ on $\mt$ was not constant, which is reflected in the $W \times B$ interaction.

\subsection{Questionnaire}
\begin{figure}[th]
    \centering
    \includegraphics[width=0.5\linewidth]{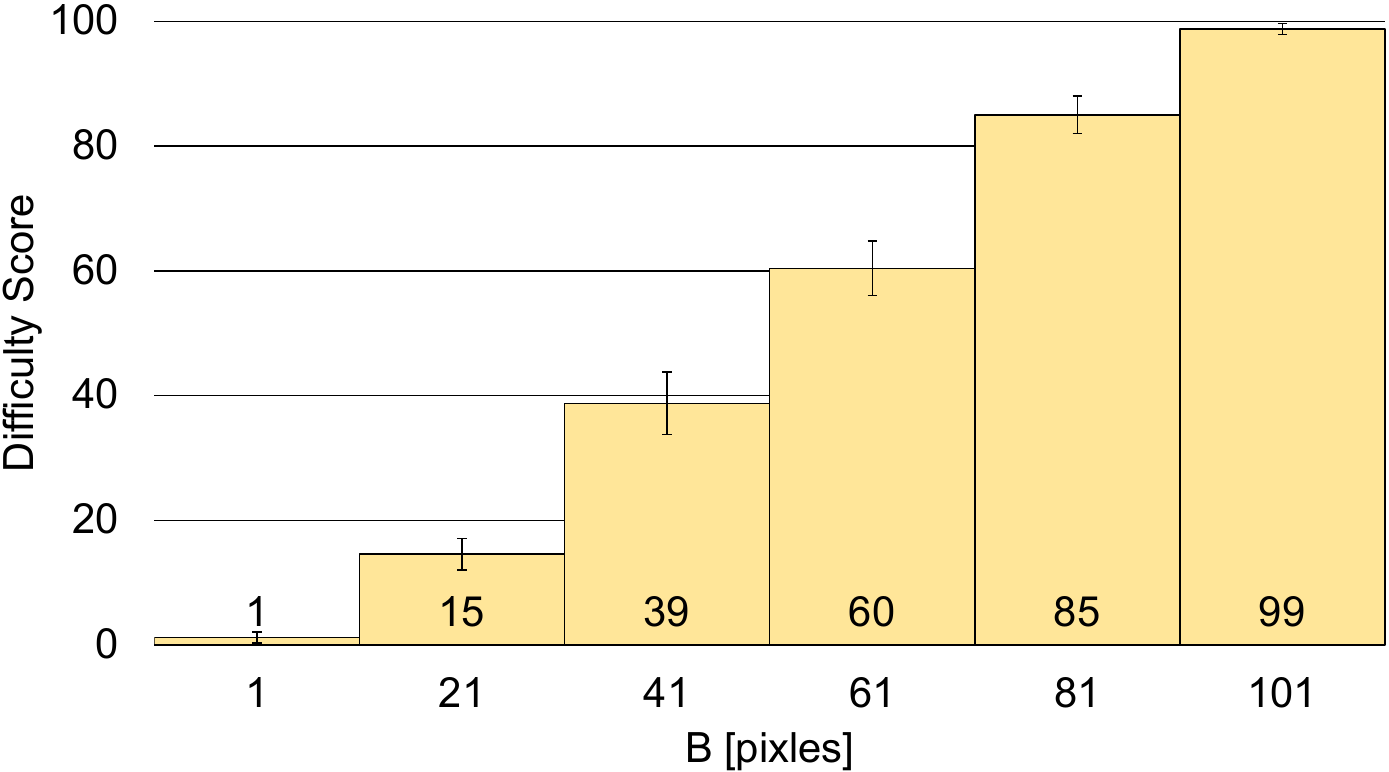}
    \caption{Subjective difficulty ratings in Experiment 1.}
    \label{fig:e1nan}
\end{figure}

\begin{figure*}[th]
    \centering
    \includegraphics[width=\linewidth]{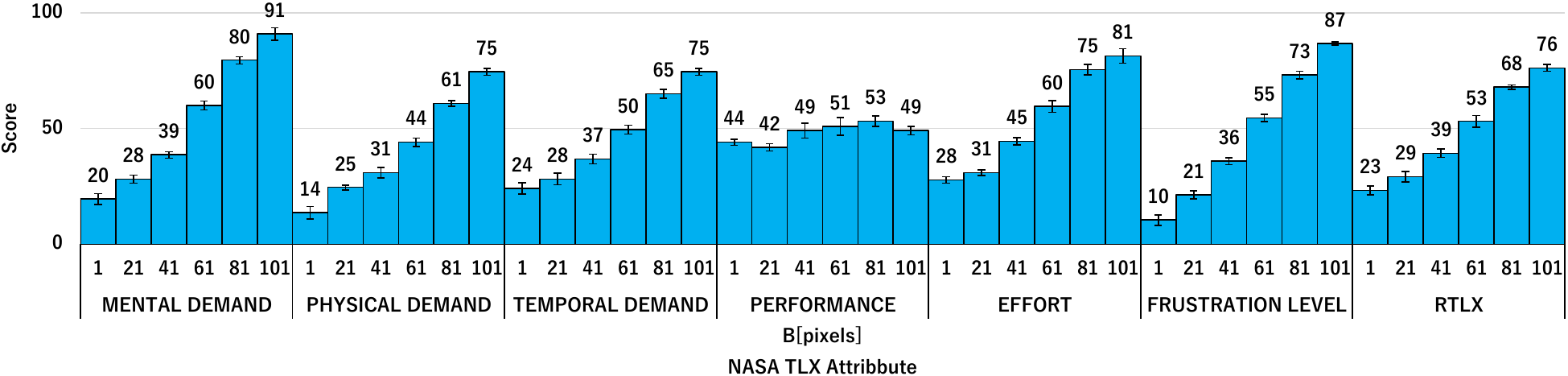}
    \caption{The six NASA-TLX items and the mean raw-TLX score at each blur level in Experiment 1.}
    \label{fig:e1tlx}
\end{figure*}

Figure~\ref{fig:e1nan} shows the subjective difficulty ratings for each $B$ condition, scored from 0 to 100.
Participants rated the task as more difficult when $B$ was larger.
Figure~\ref{fig:e1tlx} shows the six NASA-TLX items and the raw-TLX score for each $B$.
Except for the performance item, the scores tended to increase as $B$ increased.
These results are consistent with the quantitative findings that both $\er$ and $\mt$ increased as $B$ became larger, supporting the interpretation that participants subjectively perceived the negative impact of blur.


\subsection{Model Fit}
We evaluated the fit of the baseline Fitts' law (One-Part Model, Equation~\ref{eq:fitts}) and the Two-Part Model (Equation~\ref{eq:2part}).
Their adjusted $R^2$ values were 0.8342 and 0.8298, respectively; see the results for ``All 48 fitting points'' (i.e., $2_A \times 4_W \times 6_B$) in rows 1 and 5 of Table~\ref{table:e1modelAll}.
Because it has been pointed out that discussing model fit solely on the basis of $R^2$ is inappropriate \citep{Gori18perils}, we conduct a relative comparison with the refined models in Section~\ref{sec:ModelRefine}.
Nevertheless, both values were lower than the levels often reported in Fitts' law studies ($R^2 > 0.90$ \citep{Soukoreff04,Sharif20,Bertucco13,Friedlander98}), which motivated us to derive a more accurate model.

\begin{table}[th]
    \centering
    \caption{Estimated constants, model fit, and prediction accuracy in Experiment 1.}
    \label{table:e1modelAll}
    \resizebox{1.0\textwidth}{!}{
    \begin{tabular}{lll | lllll | cc | cc}
        & & & \multicolumn{7}{c|}{All 48 fitting points} & \multicolumn{2}{c}{LOOCV} \\
        \cline{4-12}
        Model & Eq. & $\mt$ prediction formulation & $a$ & $b$ & $c$ & $d$ & $e$ & adj. $R^2$ & $\aic$ & $R^2$ & $\mae$ \\
        \hline
        One-Part &\ref{eq:fitts}& $a+b\log_2{\left(\frac{A}{W}+1\right)}$ & $28.6$ & $237$ && & & $0.8342$ & $586.5$ & 0.8235 & 84.17\\
         &\ref{eq:fittsB}& $a+b\log_2{\left(\frac{A}{W}+1\right)}+c(B-1)$ & $-94.8$ & $237$ & $2.52$ & && $0.9455$ & $534.0$ & 0.9398 & 48.56\\
         &\ref{eq:fittsBB}& $a+b\log_2{\left[\frac{A}{W-c(B-1)}+1\right]}$ & $151$ & $187$ & $0.0946$ & && $0.9444$ & $535.0$ & 0.9406 & 49.16\\
         &\ref{eq:fittsAB}& $a+b\log_2{\left[\frac{A+d(B-1)}{W-c(B-1)}+1\right]}$ & $56.8$ & $200$ & $0.0738$ & $ 1.88$ && $0.9567$ & $523.9$ & 0.9487 & 44.74\\
         \hline
        Two-Part &\ref{eq:2part}& $a+b\log_2{(A)}-c\log_2{(W)}$ & $572$ & $170$ & $223$ && & $0.8298$ & $588.7$ & 0.8146 & 85.38\\
         &\ref{eq:twoPartC}& $a+b\log_2{(A)}-c\log_2{(W)}+d(B-1)$ & $446$ & $170$ & $223$ &$2.52$& & $0.9435$ & $536.7$ & 0.9358 & 50.50\\
         &\ref{eq:twoPartCD}& $a+b\log_2{(A)}-c\log_2{[W-d(B-1)]}$ & $252$ & $170$ & $172$ & $0.0977$ && $0.9414$ & $538.4$ & 0.9360 & 50.60\\
         &\ref{eq:twoPartCDE}& $a+b\log_2{[A+c(B-1)]}-d\log_2{[W-e(B-1)]}$ & $-31.6$ & $204$ & $1.67$ & $153$ &$0.0812$& $0.9543$ & $527.4$ & 0.9451 & 46.98\\
        \hline
    \end{tabular}
    }
\end{table}

\subsection{Discussion}
The overall $\er$ was high at 19.7\%, but it was 4.4\% in the no-blur condition ($B=1$).
This matches the 4--5\% $\er$ reported in prior work for appropriately balanced speed--accuracy tradeoffs \citep{MacKenzie92,Soukoreff04}.
This supports that our participants followed the instructions and attempted to operate quickly while controlling errors.
Furthermore, the effect of $B$ on $\er$ became larger as $W$ became smaller (Figure~\ref{fig:e1ER_wsp}), indicating that when the target is small, the negative impact of difficulty in visually judging whether the cursor is on the target becomes stronger.

For $\mt$, in addition to the effects of $A$ and $W$ that are commonly reported in Fitts' law research, $B$ also had a significant effect.
One plausible reason is that stronger blur weakens the visual feedback for cursor movement, and therefore the loop of submovements required to reach the target \citep{Crossman83,Meyer88} takes longer than under normal conditions.
It is also plausible that participants could not immediately determine visually that the cursor had reached the target and therefore waited until they were confident that the cursor was close enough to the center before clicking, which increased $\mt$.

In the subjective difficulty ratings, participants reported greater difficulty as $B$ increased.
In addition, the NASA-TLX scores showed a monotonic increase with $B$ for all items except performance.
In the free-description questionnaire, many participants specifically stated that stronger blur reduced their confidence that they could successfully click the target, making the task feel more difficult.
Participants also stated that when blur was strong, they progressed through the task while thinking that making mistakes was natural, and that errors were unavoidable because they could not be confident that the cursor was on the target.
Together, these findings indicate that stronger blur not only increased objective workload and difficulty but was also clearly perceived as such by the participants.

\section{Model Refinement and Application Example}
\label{sec:ModelRefine}
\subsection{Candidate Revised Models}
Neither the One- nor the Two-Part Models considers the effect of $B$ on $\mt$.
However, as shown in Figure~\ref{fig:e1MT}, $\mt$ increased by 34\%, from 838 ms ($B=1$) to 1{,}125 ms ($B=101$).
Therefore, an accurate model should incorporate $B$.

The simplest approach is to add an extra term linearly to Fitts' law, as has been done in prior studies for factors such as movement direction \citep{Murata01,Cha13}, the size of 3D objects \citep{Deng19}, and finger width \citep{Hoffmann95}.
Following that approach, adding a $B$ term to the One-Part Model yields the following equation.
\begin{equation}
    \mt =  a + b \log_2{\left(\frac{A}{W}+1\right)}+c(B-1)
    \label{eq:fittsB}
\end{equation}
Here, we subtract 1 from $B$ so that the model is consistent with the One-Part Model when $B = 1$.

\begin{figure}[th]
    \centering
    \includegraphics[width=0.60\linewidth]{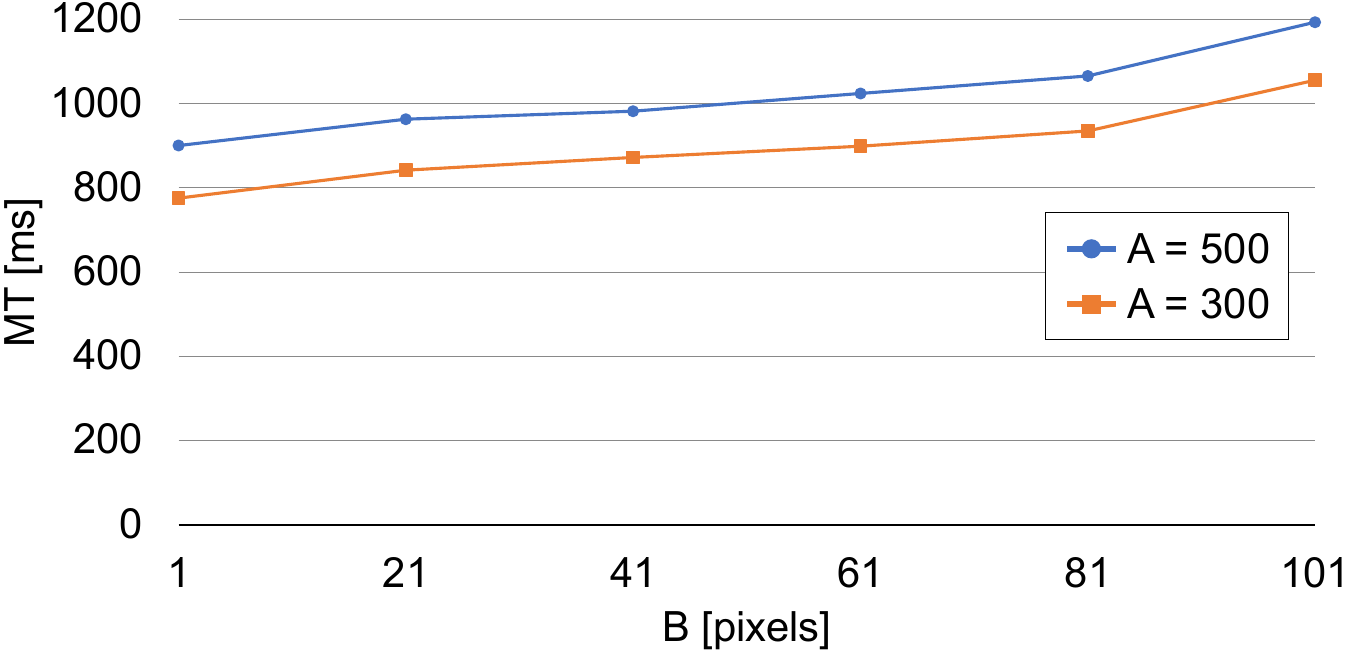}
    \caption{Effect of $B$ on $\mt$ for each $A$ in Experiment 1.}
    \label{fig:e1MT_AB}
\end{figure}

Next, we derive a model that incorporates the effect of blur on $\mt$ in a way that more closely reflects actual behavior.
Participants pointed out that when blur was strong, it became visually ambiguous whether the cursor had entered the target area of diameter $W$.
For example, in Figure~\ref{fig:studyScreen}b, when $B=41$ or higher, it is ambiguous whether the cursor is on the target, so participants may think that they must move the cursor closer to the target center to avoid failure, and this would require more time than in the $B=1$ condition.
Based on this idea, we assume that the larger $B$ becomes, the less participants can fully exploit the available target width $W$.
This leads to the following model, in which $B$ acts to reduce $W$.
\begin{equation}
    \mt =  a + b \log_2{\left[\frac{A}{W-c(B-1)}+1\right]}
    \label{eq:fittsBB}
\end{equation}
Furthermore, as shown in Figure~\ref{fig:e1MT_AB}, even when $A$ was constant, $\mt$ increased as $B$ increased.
As discussed above, this can be interpreted as the weakening of visual feedback while the cursor moves over distance $A$, which makes the feedback loop required in pointing tasks \citep{Crossman83,Meyer88} take longer.
To incorporate this effect, we further modify the model so that larger $B$ has an effect analogous to increasing $A$.
\begin{equation}
    \mt =  a + b \log_2{\left[\frac{A+d(B-1)}{W-c(B-1)}+1\right]}
    \label{eq:fittsAB}
\end{equation}

We constructed corresponding variants for the Two-Part Model as follows.
\begin{align}
\mt &= a+b\log_2{(A)}-c\log_2{(W)}+d(B-1) \label{eq:twoPartC} \\
\mt &= a+b\log_2{(A)}-c\log_2{[W-d(B-1)]} \label{eq:twoPartCD} \\
\mt &= a+b\log_2{[A+c(B-1)]}-d\log_2{[W-e(B-1)]} \label{eq:twoPartCDE}
\end{align}
Table~\ref{table:e1modelAll} presents the model fits under ``All 48 fitting points.''
In addition to adjusted $R^2$ (higher is better), we statistically compare models using the Akaike information criterion $\aic$ \citep{akaike1998information}.
Both measures impose a penalty as the number of free parameters increases.
A lower $\aic$ indicates a better and more explanatory model.
Rules of thumb for $\aic$ are as follows \citep{Burnham2003}: (1) if a candidate model's difference from the best model, $\Delta_\aic$, is less than 2, there is evidence supporting that candidate model, (2) a $\Delta_\aic$ of 2--4 indicates considerable support, (3) a $\Delta_\aic$ of 4--7 indicates much less support, and (4) a $\Delta_\aic$ greater than 10 indicates no support.

According to Table~\ref{table:e1modelAll}, within the One-Part Models, Equation~\ref{eq:fittsAB}, in which $B$ affects both $A$ and $W$, showed the highest adjusted $R^2$ and the lowest $\aic$.
The next lowest $\aic$ among the One-Part variants was the linear-addition model (Equation~\ref{eq:fittsB}), and the difference from Equation~\ref{eq:fittsAB} was $\Delta_\aic=534.0-523.9=10.1$.
Therefore, among the One-Part Models, Equation~\ref{eq:fittsAB} is the best in terms of balancing the number of free parameters and the accuracy of $\mt$ estimation.

The second-lowest $\aic$ overall was obtained by Equation~\ref{eq:twoPartCDE}, the Two-Part Model with the largest number of free parameters.
However, probably because of the penalty for additional free parameters, its adjusted $R^2$ was $0.9567-0.9543=0.0024$ lower than that of the best model, Equation~\ref{eq:fittsAB}, and its $\aic$ was $527.4-523.9=3.5$ higher.
Even considering that fewer free parameters reduce the risk of overfitting to known data, we conclude that Equation~\ref{eq:fittsAB} is still the best model.

We further compared the accuracy of $\mt$ estimation for unseen $A\times W\times B$ conditions using leave-one-($A$, $W$, $B$)-out cross-validation (LOOCV).
Table~\ref{table:e1modelAll} reports the resulting $R^2$ and mean absolute error $\mae$ in milliseconds in the two rightmost columns.
Higher $R^2$ and lower $\mae$ indicate better accuracy.
The results were consistent with the fits obtained from all 48 data points, and Equation~\ref{eq:fittsAB} again performed best.
This alleviates the concern that Equation~\ref{eq:fittsAB} overfits the known $\mt$ data, and it indicates that this equation can estimate $\mt$ most accurately even for unseen task conditions.

\subsection{An Application of the Revised Model}
One example of what cannot be achieved with conventional Fitts' law is to adjust $W$ so that $\mt$ remains stable even when the blur level changes.
If that were possible, users could, for example, point comfortably even when a projector cannot be placed within its effective distance range, or when visual acuity declines, while maintaining $\mt$s similar to those under normal vision.
To realize such ideas, an additional challenge would be to measure the amount of blur caused by projector defocus or reduced visual acuity.
Here, however, we assume that such measurement can somehow be achieved and discuss the applicability of the model under that assumption.

Suppose that under a condition corresponding to $B=101$, a user must select a target with $A=300$ and $W=18$ pixels.
Let us assume that to obtain the same $\mt$ as in the $B=1$ condition, the target size must be increased by $\Delta W$ pixels.
Using the regression constants obtained in Experiment 1, the best model (Equation~\ref{eq:fittsAB}) predicts that the following equality must hold for the $\mt$ values computed at $B=1$ and $B=101$ to be the same.
\begin{equation}
\label{eq:egB}
\resizebox{0.88\linewidth}{!}{$
56.8+200 \log_2{\left[\frac{300+1.88(1-1)}{18-0.0738(1-1)} + 1\right]} = 56.8+200 \log_2{\left[\frac{300+1.88(101-1)}{18-0.0738(101-1)+\Delta W}+1\right]}
$}
\end{equation}
Solving this equation for $\Delta W$ gives 18.66.
Thus, increasing $W$ by 19 pixels to 37 pixels is expected to produce $\mt$s comparable to those obtained when the cursor and target are clearly visible.
Generalizing this calculation, the required $\Delta W$ to obtain the same $\mt$ as in the no-blur condition for a given $B$ can be expressed as follows.
\begin{equation}
\label{eq:egfittsAB}
a+b \log_2{\left[\frac{A+d(1-1)}{W-c(1-1)} + 1\right]} = a+b \log_2{\left[\frac{A+d(B-1)}{W-c(B-1)+\Delta W}+1\right]}
\end{equation}
Solving this for $\Delta W$ yields the following equation.
\begin{equation}
    \label{eq:egfittsAB2}
    \Delta W=\frac{(B-1)(cA+dW)}{A}
\end{equation}

\section{Experiment 2: A Follow-up on Target-Size Adjustment Application}
\subsection{Overview}
We proposed an $\mt$ prediction model that incorporates blur level $B$ (Equation~\ref{eq:fittsAB}) and a method for adjusting $W$ to prevent blur-induced increases in $\mt$ (Equation~\ref{eq:egfittsAB2}).
For the underlying model (Equation~\ref{eq:fittsAB}), cross-validation has already shown that it can estimate $\mt$ under new $B$ and $W$ conditions with high accuracy.
Accordingly, the model predicts that the measured $\mt$ after enlarging the target from $W$ to $W+\Delta W$ should be close to the estimated value.

Experiment 2 was conducted as a follow-up to strengthen this claim further.
Specifically, we examined whether enlarging the target by $\Delta W$ would actually produce $\mt$ values close to those in the $B=1$ condition.
Because the amount by which $\mt$ increases with blur varies across individuals, as reflected in the error bars in Figure~\ref{fig:e1MT}, the required $\Delta W$ is also expected to differ by participant.
Hence, we have to fit Equation~\ref{eq:fittsAB} separately to each participant and calculate $\Delta W$ from the resulting coefficients.

Previous studies have already examined the fit of Fitts' law for individual participants \citep{Wobbrock08error,Wobbrock11dim}, and it is natural for the fit to decrease relative to analyses based on mean $\mt$ across many participants.
For example, \cite{Sharif20} reported that Pearson's $r$ averaged 0.85 to 0.87 across individual participants.
Yet, prior work does not provide a basis for deciding what level of $R^2$ would be sufficient when fitting our proposed model (Equation~\ref{eq:fittsAB}) to each participant individually.

In addition, because this follow-up experiment involved only six participants, the reliability of null-hypothesis significance testing is limited.
We thus discuss $\mt$ stability primarily by comparing the reduction in the change ratio against the no target-size correction condition, although we report significance tests following a standard analysis procedure.

This experiment was conducted in two blocks.
In the first block, participants performed the same task as in Experiment 1, and we fitted Equation~\ref{eq:fittsAB} separately for each participant.
Using those coefficients, we calculated $\Delta W$ with Equation~\ref{eq:egfittsAB2} to determine how much the target should be enlarged to obtain $\mt$ close to that in the no-blur condition.
In the second block, we actually enlarged the target by $\Delta W$ and evaluated whether $\mt$ comparable to the $B=1$ condition was observed under each $B$ condition.

We used the same task and apparatus as in Experiment 1.
Because running two blocks was expected to increase participant burden relative to Experiment 1, we reduced the number of targets per session from 21 to 15.

\subsection{Participants}
Six new participants aged 20--23 years took part in the experiment (mean age $=$ 21.0 years, SD $=$ 1.10).
All participants reported that they were right-handed and normally used a mouse with their right hand, so all participants operated the mouse with their right hand in the experiment.
They had normal or corrected-to-normal vision.
Four participants wore glasses and two used no correction.

\subsection{Design}
$A$ had two levels (300 and 1{,}100 pixels), $W$ had four levels (12, 18, 36, and 78 pixels), and $B$ had six levels (1, 21, 41, 61, 81, and 101 pixels).
The $\id$ ranged from 2.28 to 6.53 bits.
We increased the upper level of $A$ from 500 to 1{,}100 pixels.
This was done because measuring a wider range of $\id$ is expected to improve the reliability of regression constants in model fitting \citep{stat462node102,MacKenzie92}, and increasing the longest distance allowed us to achieve that.
Participants completed the $2_A \times 4_W \times 6_B$ combinations in two blocks, first without target-size correction and then with correction ($C$: with vs. without correction).

\subsection{Procedure}
Except for the fact that two blocks were conducted, the procedure was the same as in Experiment 1, including the target conditions in the practice sessions, the randomization of $B$ conditions, and the questionnaires.
The second block also included the $B=1$ condition to equalize the number of trials across the two blocks and enable a fair comparison for the two $C$ conditions.
However, $W$ is not corrected when $B=1$ is substituted into Equation~\ref{eq:egfittsAB2}, and thus it would not be essential if the sole purpose were to evaluate $\mt$ stability.
The experiment took 60 minutes per participant.

\section{Results of Experiment 2}
We analyzed the data from 8{,}640 data-collection trials obtained from six participants.
The independent variables were $A$, $W$, $B$, and $C$.

\subsection{Error Rate}
\begin{figure}[th]
    \centering
    \includegraphics[width=0.9\linewidth]{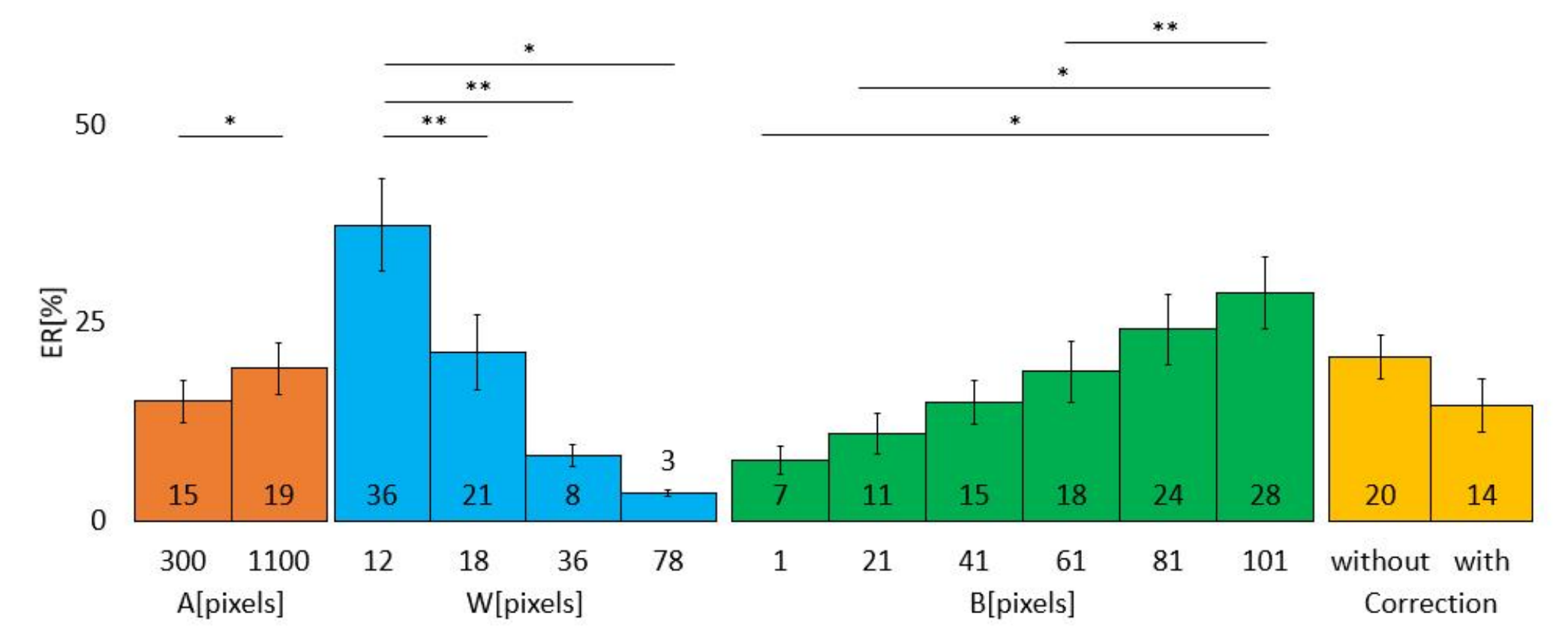}
    \caption{Effects of $A$, $W$, $B$, and $C$ on $\er$ in Experiment 2.}
    \label{fig:e2ER}
\end{figure}

\begin{figure}[th]
    \centering
    \includegraphics[width=\linewidth]{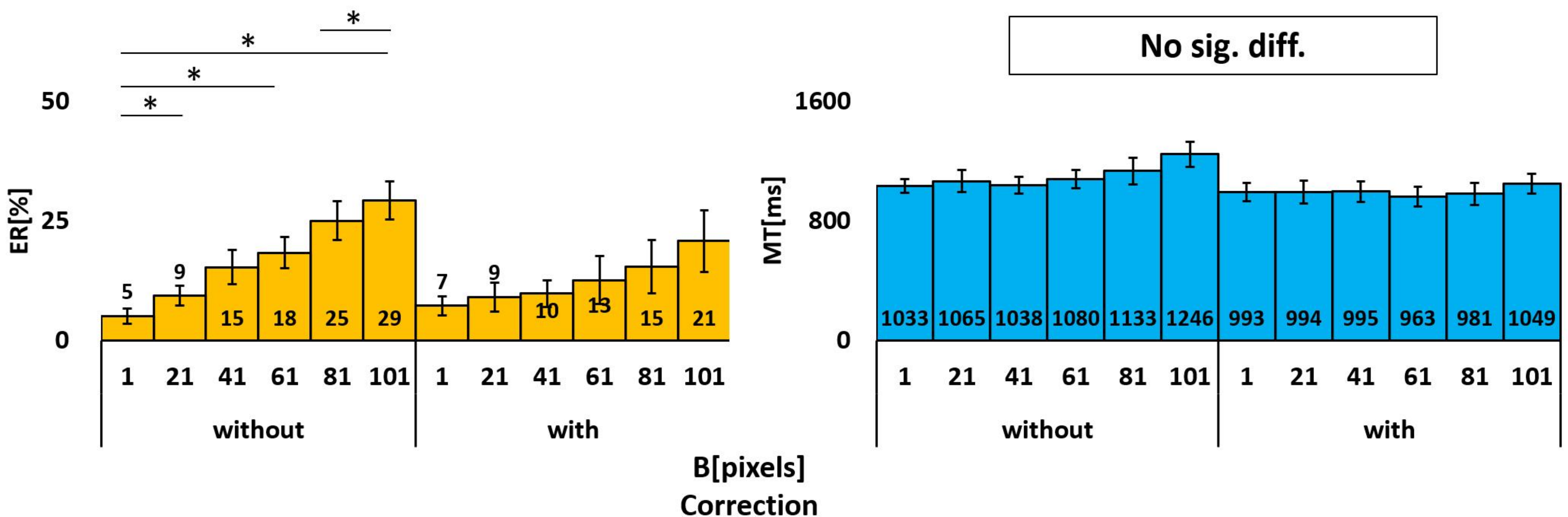}
    \caption{Effects of $B$ on $\er$ (left) and $\mt$ (right) for each $C$ condition in Experiment 2.}
    \label{fig:e2ERMT_bc}
\end{figure}

We observed 1{,}483 errors, resulting in an overall $\er$ of 17.2\%.
Figure~\ref{fig:e2ER} shows the effects of each condition on $\er$, and the left side of Table~\ref{table:e2ERMT} shows the ANOVA results.
Figure~\ref{fig:e2ERMT_bc} left further shows the effect of $B$ on $\er$ for each $C$ condition.
The effect of $B$ tended to be smaller with correction than without correction.

\begin{table}[t]
    \centering
    \caption{Results of the RM-ANOVA for $\er$ and $\mt$ in Experiment 2.}
    \label{table:e2ERMT}
    \resizebox{1\textwidth}{!}{%
    \begin{tabular}{l|lll|lll}
            &\multicolumn{3}{c|}{$\er$}&\multicolumn{3}{c}{$\mt$}\\
             \multicolumn{1}{c|}{Factors} &\multicolumn{1}{c}{$F$} &\multicolumn{1}{c}{$p$} &\multicolumn{1}{c|}{$\eta_{p}^{2}$} &\multicolumn{1}{c}{$F$} &\multicolumn{1}{c}{$p$} &\multicolumn{1}{c}{$\eta_{p}^{2}$} \\
             \hline
                $A$&$F(1,5)=9.664$&$<0.05$&0.659&$F(1,5)=353.0$&$<0.001$&0.986\\
                $W$&$F(3,15)=30.10$&$<0.001$&0.858&$F(3,15)=150.0$&$<0.001$&0.968\\
                $B$&$F(5,25)=17.88$&$<0.001$&0.781&$F(5,25)=7.639$&$<0.001$&0.604\\
                $C$&$F(1,5)=4.800$&$0.08001$&0.490&$F(1,5)=7.645$&$<0.05$&0.605\\
                \hline
                $A\times W$&$F(3,15)=2.979$&$0.06496$&0.373&$F(3,15)=4.150$&$<0.05$&0.454\\
                $A\times B$&$F(5,25)=0.9162$&$0.4866$&0.155&$F(5,25)=2.317$&$0.07355$&0.317\\
                $W\times B$&$F(15,75)=10.84$&$<0.001$&0.684&$F(15,75)=0.5485$&$0.9038$&0.099\\
                $A\times C$&$F(1,5)=0.6244$&$0.4652$&0.111&$F(1,5)=2.554$&$0.1709$&0.338\\
                $W\times C$&$F(3,15)=14.39$&$<0.001$&0.742&$F(3,15)=2.092$&$0.1442$&0.295\\
                $B\times C$&$F(5,25)=3.860$&$<0.01$&0.436&$F(5,25)=3.090$&$<0.05$&0.382\\
                \hline
                $A\times W\times C$&$F(3,15)=0.6071$&$0.6206$&0.108&$F(15,75)=0.3039$&$0.9937$&0.057\\
                $A\times W\times B$&$F(15,75)=1.024$&$0.4409$&0.170&$F(15,75)=1.958$&$0.1638$&0.281\\
                $A\times B\times C$&$F(5,25)=0.06569$&$0.9967$&0.013&$F(5,25)=0.9328$&$0.4767$&0.157\\
                $W\times B\times C$&$F(15,75)=1.669$&$0.07596$&0.250&$F(15,75)=2.191$&$<0.05$&0.305\\
                \hline
                $A\times W\times B\times C$&$F(15,75)=1.516$&$0.1212$&0.233&$F(15,75)=0.8708$&$0.5982$&0.148\\
                \hline
    \end{tabular}
    }
\end{table}

\subsection{Movement Time}
\begin{figure}[th]
    \centering
    \includegraphics[width=0.82\linewidth]{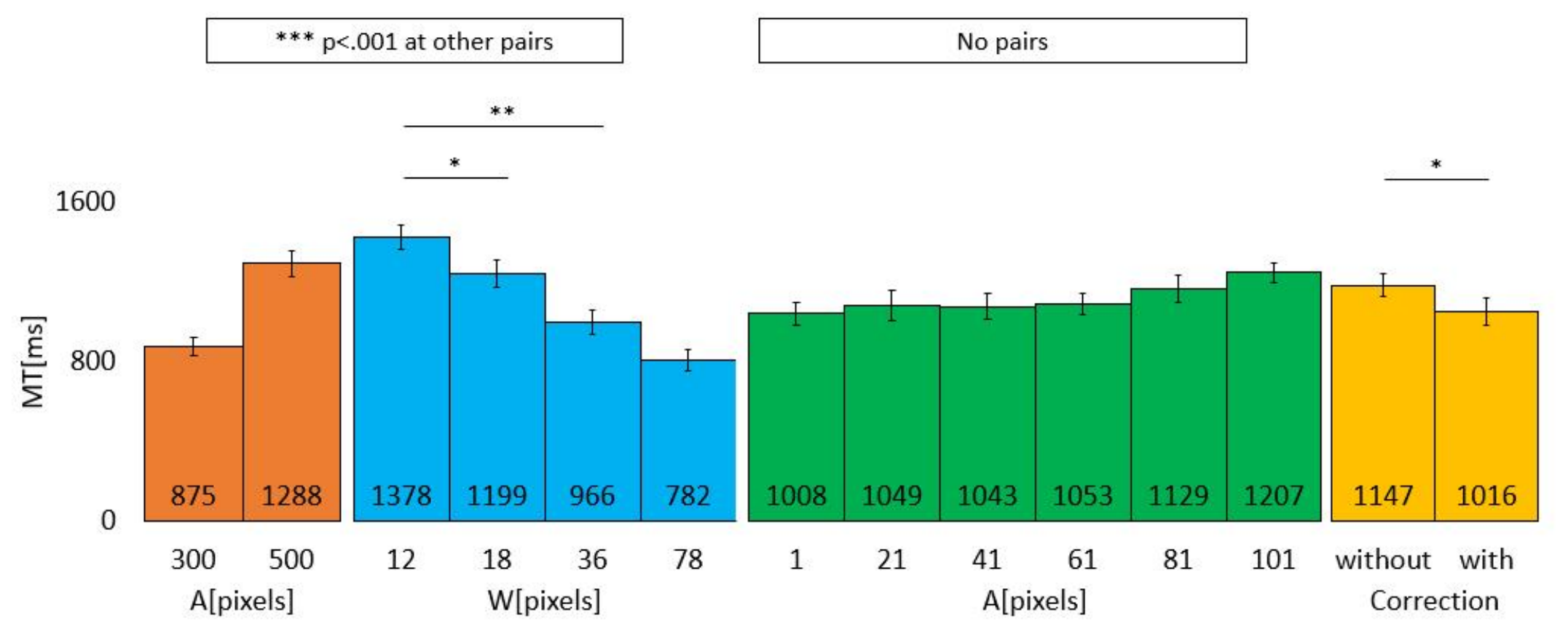}
    \caption{Effects of $A$, $W$, $B$, and $C$ on $\mt$ in Experiment 2.}
    \label{fig:e2MT}
\end{figure}

\begin{figure*}[th]
    \centering
    \includegraphics[width=\linewidth]{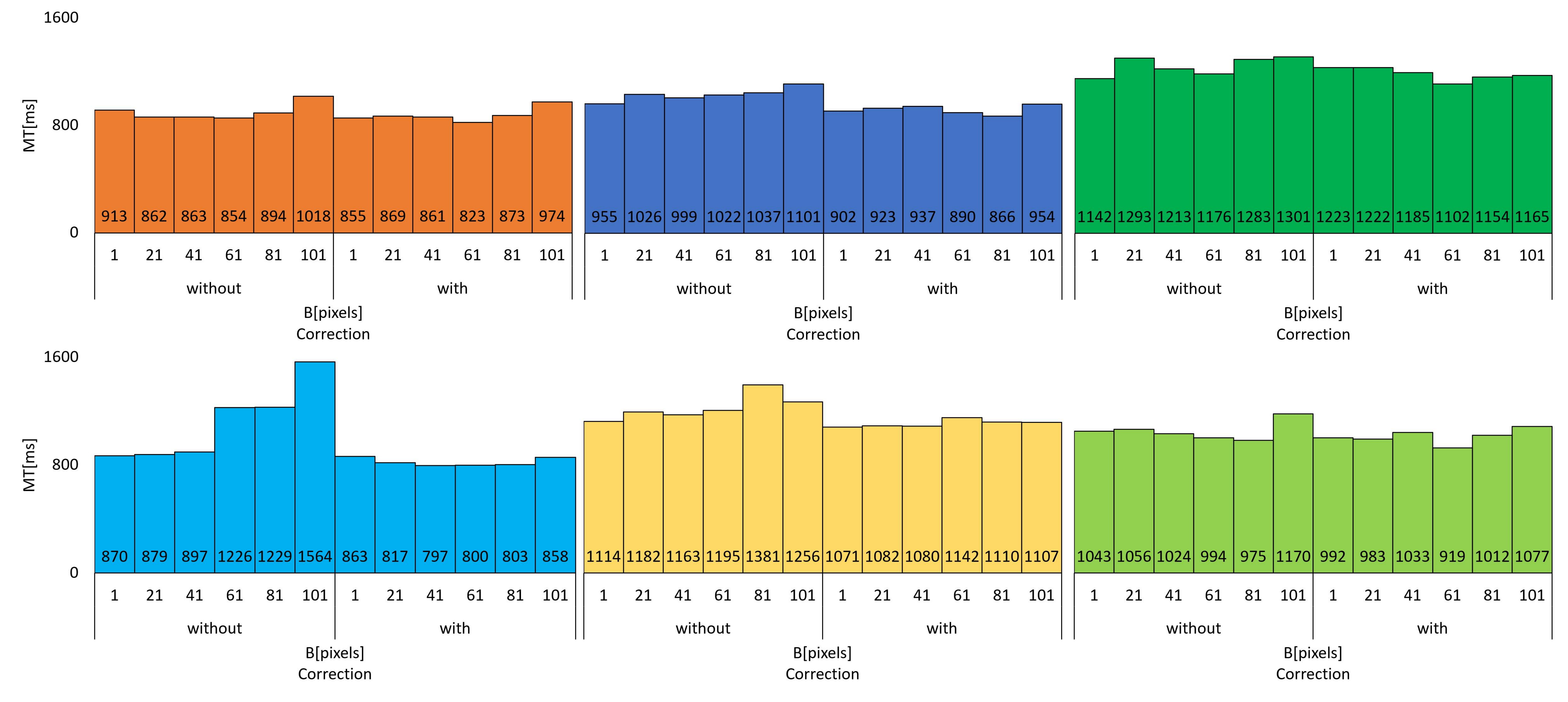}
    \caption{Effects of $B$ on $\mt$ for each $C$ condition and for each participant in Experiment 2.}
    \label{fig:e2MTp_bc}
\end{figure*}

We analyzed the data from 7{,}157 error-free trials.
The overall mean $\mt$ was 1{,}048 ms.
Figure~\ref{fig:e2MT} shows the effects of each condition on $\mt$, and the right side of Table~\ref{table:e2ERMT} shows the ANOVA results.
Figure~\ref{fig:e2ERMT_bc} right further shows the effect of $B$ on $\mt$ for each $C$ condition.
The effect of $B$ again tended to be smaller with correction than without correction.
This pattern was also observed when the data were analyzed separately for each participant (Figure~\ref{fig:e2MTp_bc}).


\subsection{Effect of Target-Size Correction}
In the no-correction condition, the baseline was $\mt=1{,}033$ ms at $B=1$, whereas $\mt=1{,}246$ ms at $B=101$, which is an increase of 213 ms (20.6\%; Figure~\ref{fig:e2ERMT_bc} right).
In contrast, in the correction condition, the baseline was $\mt=993$ ms at $B=1$, whereas $\mt=1{,}049$ ms at $B=101$, which is an increase of 56 ms (5.6\%).
Therefore, the effect of $B$ on $\mt$ was smaller when correction was applied.

According to Figure~\ref{fig:e2ERMT_bc} right, no significant differences in $\mt$ were found between $B$ conditions regardless of whether correction was applied.
However, this should not be interpreted as evidence that similar $\mt$ values can be obtained across $B$ conditions even without correction.
Because Experiment 2 involved only six participants, the sample size may simply have been too small to detect significant differences.
This interpretation is supported by Experiment 1, which had 12 participants and no correction, and in which $\mt$ significantly increased as $B$ increased, from $\mt=838$ ms at $B=1$ to $\mt=1{,}125$ ms at $B=101$ (a 34.2\% increase, Figure~\ref{fig:e1MT}).

Because the participants differed between the two experiments, direct comparison is difficult due to individual differences.
Nevertheless, it is noteworthy that in Experiment 2, target-size correction reduced the effect of $B$ on $\mt$ to 5.6\%.
In addition, the standard deviation of $\mt$ across the six $B$ conditions was 80.45 ms without correction and 28.76 ms with correction, meaning that correction reduced the variation in $\mt$ by 64\%.
These results support the effectiveness of target-size correction in bringing $\mt$ closer across $B$ conditions.

\subsection{A Quantitative Evaluation for MT Stabilization by Correction}
Probably because the number of participants was small, the results did not match our expectations as explained below, but we report the results of the two one-sided tests (TOST) procedure as guidance for future studies.
The hypothesis here was that, in the correction condition, $\mt$ would remain equivalent to that in the $B=1$ condition even if $B$ increased.

First, for each participant's $\mt$ values in the second block, we computed the differences between the $B=1$ condition and each of the $B=21$, 41, 61, 81, and 101 conditions within each $A\times W$ condition.
We then examined whether the mean difference fell within a predefined equivalence bound.
The equivalence bound was defined using a small effect size, $dz=0.2$ \citep{Cohen88}.
Specifically, for each comparison, we multiplied the standard deviation of the differences across the six participants by 0.2 and used that value as the equivalence bound $\Delta$.
This criterion means that differences smaller than a small effect size are regarded as practically negligible.
The significance level was set to $\alpha=0.05$.
In TOST, two one-sided tests are conducted, namely whether the mean difference is greater than $-\Delta$ and whether it is smaller than $+\Delta$, and equivalence is concluded only if both tests are significant.

We conducted a total of 40 equivalence tests by comparing $B=1$ with the other five $B$ levels for each of the eight ($2_A \times 4_W$) conditions.
To control the familywise error rate under multiple comparisons, we applied Holm correction to the $p$-values.
As a result, equivalence was not established in any comparison.
Therefore, we did not obtain statistical evidence that the $\mt$s under $B \ne 1$ conditions were equivalent to those in the $B=1$ condition.

\subsection{Questionnaire}
\begin{figure}[th]
    \centering
    \includegraphics[width=0.63\linewidth]{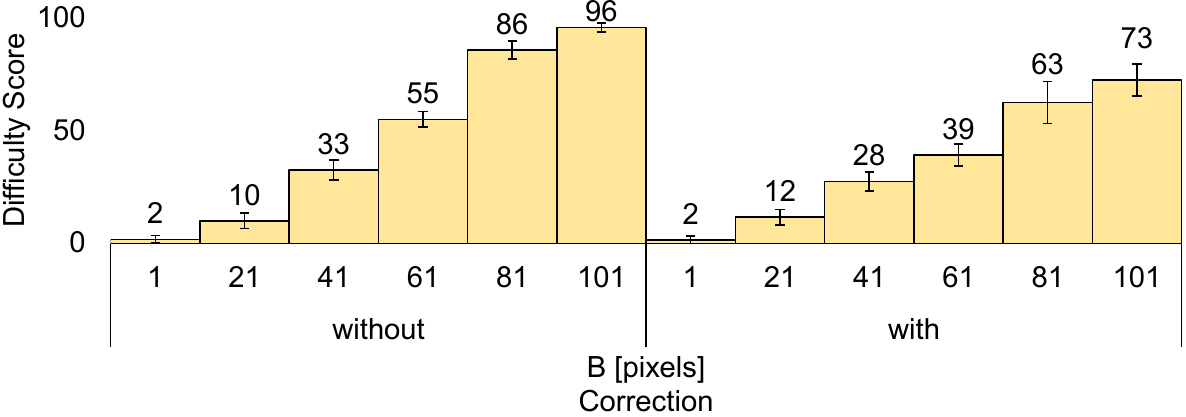}
    \caption{Subjective difficulty ratings in Experiment 2.}
    \label{fig:e2diff}
\end{figure}

\begin{figure*}[th]
    \centering
    \includegraphics[width=\linewidth]{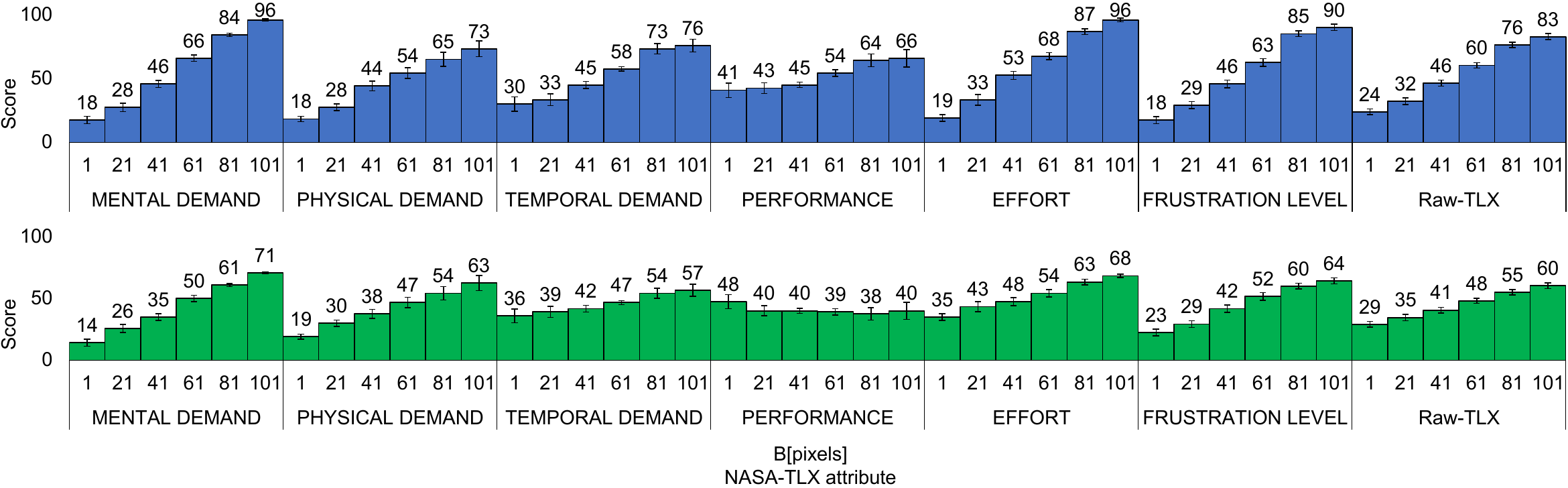}
    \caption{The mean raw-TLX scores at each blur level in Experiment 2, where the slopes are smaller with correction (bottom) than without correction (top).}
    \label{fig:e2TLX}
\end{figure*}

Figure~\ref{fig:e2diff} shows that the participants again rated the task as more difficult as $B$ increased, but the perceived difficulty tended to decrease when correction was applied.
Figure~\ref{fig:e2TLX} shows the raw-TLX scores for each $B$.
With correction (Figure~\ref{fig:e2TLX}, bottom), the slopes tended to be smaller than without correction (Figure~\ref{fig:e2TLX}, top).
Thus, in the target-size correction condition, the increase in raw-TLX with increasing $B$ was suppressed.
This indicates that participants also subjectively perceived the beneficial effect of enlarging target size.

\subsection{Model Fit}
To keep $\mt$ stable by enlarging target size, $\Delta W$ must be computed appropriately, which requires the proposed model to fit each participant's $\mt$ data in the first block (i.e., no-correction condition) well.
Otherwise, enlarging the target by $\Delta W$ might still be insufficient, or the target might become unnecessarily large, which would increase screen occupancy even if $\mt$ were reduced.

Table~\ref{table:e2model} (right) shows the fit of the proposed model to each participant's $\mt$ data.
The adjusted $R^2$ values ranged from 0.8029 to 0.9135.
As noted above, we cannot determine from the present experiment alone whether these values are high or low, but they can serve as benchmarks for comparison in future replications.
What our results show is that, at least with fit in this range, it was possible to compute $\Delta W$ that reduced the blur-induced increase in $\mt$, suggesting that an adjusted $R^2$ of around 0.8 may be necessary.

Participant 6 had $d=0$, indicating that blur did not have an effect analogous to increasing $A$.
This suggests that the blur-induced increase in $\mt$ was sufficiently explained by the term $W-c(B-1)$, and that this participant spent a large amount of time on fine cursor adjustment near the target.

The results of Experiment 2 can also be used to examine the reproducibility of Experiment 1.
Using the no-correction condition in Experiment 2, we evaluated the fit of the baseline model (the One-Part Model of Fitts' law, Equation~\ref{eq:fitts}) and the proposed model that performed best in Experiment 1 (Equation~\ref{eq:fittsAB}) based on the mean $\mt$ across all participants (Table~\ref{table:e2model} bottom).
The proposed model showed a higher adjusted $R^2$ than the baseline (0.9622 vs. 0.8581).
Because the $\aic$ difference was greater than 10 ($605.9-544.3=61.6$), the proposed model was supported as the more accurate estimator of $\mt$.

The participant-wise fits also consistently favored the proposed model, which yielded higher adjusted $R^2$ and lower $\aic$ for every participant.
Only Participants 3 and 6 showed an $\aic$ difference of 2 or less between the baseline and proposed models, whereas the other four participants showed differences greater than 2.
This again indicates that the baseline model is not supported.

\begin{table}[t]
    \centering
    \caption{Model fit in the no-correction condition of Experiment 2.}
    \label{table:e2model}
    \resizebox{1.0\textwidth}{!}{%
    \begin{tabular}{c|cc|cc|cccc|cc}
        & \multicolumn{4}{c|}{Baseline: $\mt=a+b\log_2{\left(\frac{A}{W}+1\right)}$}&\multicolumn{6}{c}{Proposed model: $\mt=a+b\log_2{\left[\frac{A+d(B-1)}{W-c(B-1)}+1\right]}$}\\
        \cline{2-11}
        Participant & $a$ & $b$ & $\textit{adj.}R^2$ & $\aic$ & $a$ & $b$ & $c$ & $d$ & $\textit{adj.}R^2$ & $\aic$\\
        \hline
        1 & $55.87$ & $199.3$ & $0.8962$ & $569.2$ & $77.29$ & $185.1$ & $0.05711$ & $0.1201$ & $0.9135$ & $562.3$\\
        2 & $64.21$ & $230.9$ & $0.8049$ & $618.7$ & $-80.82$ & $242.8$ & $0.01945$ & $2.739$ & $0.8498$ & $608.0$\\
        3 & $245.1$ & $236.7$ & $0.8532$ & $604.7$ & $207.1$ & $233.9$ & $0.03121$ & $0.8481$ & $0.8643$ & $602.8$\\
        4 & $-261.8$ & $318.0$ & $0.4628$ & $723.7$ & $-413.6$ & $279.1$ & $0.1173$ & $4.914$ & $0.8970$ & $646.2$\\
        5 & $156.1$ & $251.3$ & $0.7586$ & $639.8$ & $75.92$ & $247.1$ & $0.05095$ & $1.787$ & $0.8029$ & $631.9$\\
        6 & $68.07$ & $231.0$ & $0.8957$ & $583.7$ & $92.97$ & $217.4$ & $0.04679$ & $0$ & $0.9030$ & $582.0$\\
        \hline
        Mean $\mt$ over all participants & $54.59$ & $244.5$ & $0.8581$ & $605.9$ & $80.08$ & $215.6$ & $0.09162$ & $0.6946$ & $0.9622$ & $544.3$\\
        \hline
    \end{tabular}
    }
\end{table}

\section{General Discussion}
\subsection{Findings on the Effect of Blur on Pointing Performance}
In experiments in which the visual clarity of the entire screen was reduced by Gaussian blur, we showed that increasing $B$ worsened both $\er$ and $\mt$.
For $\mt$, which is the primary focus of this study, Experiment 1 showed that the strongest blur condition increased $\mt$ by 34.2\% relative to the no-blur condition.
This clearly demonstrates the negative effect of reduced visual clarity on pointing performance.

Participants reported that blur made the boundaries of the cursor and target unclear, making it difficult to judge with confidence whether the cursor was inside the target area.
As a result, participants faced a tradeoff between delaying the click and operating cautiously, for example by moving the cursor around the target and pressing the mouse button when the red color of the target was occluded, and clicking without confidence and thereby causing errors.
Furthermore, the NASA-TLX and subjective-difficulty scores indicated that participants experienced greater workload, mental fatigue, and operational difficulty as blur became stronger.

\subsection{Model Fit}
The results of Experiments 1 and 2 indicate that the best model incorporating $B$, namely Equation~\ref{eq:fittsAB}, estimated $\mt$ more accurately than conventional Fitts' law (Equation~\ref{eq:fitts}).
We also compared a simple linear-addition model (Equation~\ref{eq:fittsB}), as has been done in several previous studies \citep{Murata01,Cha13,Deng19,Hoffmann95}.
However, its $\aic$ was more than 10 points larger than that of Equation~\ref{eq:fittsAB}, indicating that the effect of $B$ needs to be incorporated in a way that reflects user behavior more realistically.

Because $\mt$ increased with $B$ even when $A$ was held constant (Figure~\ref{fig:e1MT_AB}), it is reasonable that Equation~\ref{eq:fittsAB} outperformed Equation~\ref{eq:fittsBB}, which assumes only a reduction in $W$.
We further conducted LOOCV to test whether each model could accurately estimate $\mt$ under unseen target conditions, and Equation~\ref{eq:fittsAB} remained the best.
Note that separating the $A$ and $W$ terms (i.e., using the Two-Part Models) did not improve fit enough to outperform the One-Part Models.

In Experiment 2, target-size correction required fitting the model to each participant's $\mt$ data individually, and the resulting adjusted $R^2$ ranged from 0.8029 to 0.9135.
As noted above, it is natural for the fit of Fitts' law to be low when individual-participant data are used \citep{Sharif20}.
Moreover, within-participant $\mt$ data are likely to contain additional noise, such as order effects, for example whether the most difficult condition ($B=101$, $W=12$) happened to be performed late in the experiment after the participant had become accustomed to the task.
In general, balancing experimental conditions and averaging $\mt$ across participants help prevent such noise from contributing directly to model fit, and therefore the range $0.8029 \leq \mathrm{adj.\,}R^2 \leq 0.9135$ should not be regarded as inferior to previously proposed Fitts' law models.
Because there is no consensus in HCI regarding what threshold should be considered sufficiently high for fit to individual data, the fact that target-size correction worked reasonably well even at around an adjusted $R^2$ of 0.8 can serve as a useful benchmark.
Taken together, Equation~\ref{eq:fittsAB} is practical enough to serve as a basis for adapting UI design to individual users' motor abilities and visual characteristics, as well as to device and environmental factors.

\subsection{Benefits of Target-Size Correction}
Experiment 2 showed that the $\Delta W$ derived from the proposed model reduced the increase in $\mt$ associated with increasing blur relative to the no-correction condition.
The idea behind this correction is to compensate for the fact that blur makes the target area less clearly distinguishable and therefore prevents users from fully exploiting the available width $W$ by physically enlarging the target on the UI side.
The effectiveness of this correction was also reflected in subjective difficulty and NASA-TLX, showing that both performance measures and experience measures improved.

More quantitatively, whereas the no-correction condition showed a 20.6\% increase in $\mt$ at the maximum blur level relative to the baseline, the correction condition reduced this increase to 5.6\%.
This means that enlarging the target lowered task difficulty and offset the cognitive and motor delays caused by blur.
In addition, the standard deviation of $\mt$ across the six $B$ conditions decreased by about 64\%, from 80.45 ms without correction to 28.76 ms with correction.
This demonstrates a stabilizing effect on $\mt$ even when blur level changed.

On the other hand, for some participants, $\mt$ was not completely flattened even in the correction condition (Figure~\ref{fig:e2MTp_bc}), which is one limitation of the proposed method.
Moreover, enlarging targets increases their screen occupancy and may create new tradeoffs, such as reducing the information density that can be displayed on one screen or occluding neighboring targets.
Previous studies have discussed various target-enlargement techniques and their adverse effects \citep{Zhang24,McGuffin05,Zhai03online,Cockburn06online,Gutwin02}, and similar issues may arise in our approach as well.

\subsection{Implications and Directions for Future Research}
In prior research on pointing, it has been customary to state that only participants with normal or corrected-to-normal vision were included.
Our results support the necessity of such restrictions when researchers seek strict experimental control, because people who perceive the screen as strongly blurred will likely exhibit performance that differs substantially from that of others and may introduce noise into experiments comparing devices or validating performance models.
Of course, recruiting a diverse participant pool that includes such effects is also meaningful, but if researchers do not want that from the perspective of experimental control, they may reasonably consider appropriate restrictions.

The best model in this study, Equation~\ref{eq:fittsAB}, introduces blur as a continuous parameter $B$, which may make it easier to compare the degree of reduced visual feedback on a common scale across device factors, environmental factors, and user factors.
However, this study evaluated only device-related blur.
Testing whether the same model can also estimate $\mt$ accurately for blur arising from environmental factors, such as glare, and user factors, such as eye fatigue, remains an important direction for future work.

From the perspectives of accessibility and universal design, the findings may also be applicable to a wide range of real-world UI design problems.
For example, the findings could support OS-level accessibility features that automatically adjust UI element size according to the visual acuity of users whose vision has declined because of aging or disease.
Relatedly, \cite{w3cContrastMinimum} provides guidance for readable visual presentation for users with low vision, but our results suggest that low vision can also negatively affect UI operation.
Thus, appropriate UI design guidelines for interactive tasks should also be developed in the future.

\subsection{Limitations}
In our experimental system, blur level $B$ was the kernel size of Gaussian blur, together with the corresponding blur strength $\sigma$, and this does not perfectly reproduce the blur caused by reduced visual acuity or projector defocus, which is a limitation in terms of external validity.
Another limitation is that our participants were all young adults.
Furthermore, the input device was restricted to a mouse, and actual UIs may employ mouse-over effects, such as changes in color or highlighting.
Future work should test whether the effects of blur level and model fit generalize to a wider age range, other input devices like touchscreens and VR controllers, and extra visual effects.

Experiment 2 in particular needs replication.
No significant differences in $\mt$ were found among the $B$ conditions regardless of whether target-size correction was applied (Figure~\ref{fig:e2ERMT_bc}).
Because the number of participants was small, the results were more susceptible to low statistical power, and therefore the reliability of the conclusions drawn from Experiment 2 particularly depends on replication.

Although we evaluated the usefulness of correcting target size based on $B$, the proposed model also makes it possible to prevent performance degradation by correcting $A$, for example by changing the target layout.
If only $A$ is adjusted to obtain the same $\mt$ as at $B=1$, and if the corrected target distance is defined as $A'=A-\Delta A$, then $\Delta A=(B-1)\left(d+\frac{cA}{W}\right)$.
If both $A$ and $W$ are adjusted, and the corrected target distance and size are defined as $A'$ and $W'$, respectively, then they only need to satisfy $\frac{A}{W}=\frac{A'+d(B-1)}{W'-c(B-1)}$, which yields $\Delta W=-\frac{W}{A}\Delta A+\frac{(B-1)(cA+dW)}{A}$.
Therefore, once either $\Delta A$ or $\Delta W$ is determined, the other is calculated automatically.
In Experiment 2, we evaluated only the stabilization of $\mt$ through adjustment of $W$, but future work should also test the effectiveness of adjusting $A$ in this manner.
Ultimately, we would like to implement a UI design support tool that can consider constraints such as enlarging $W$ as much as possible without interfering with neighboring targets and then adjusting layout through $A$ to stabilize $\mt$.

\section{Conclusion}
In this study, we investigated how the strength of a blur effect applied to the screen affects pointing performance.
The results showed that stronger blur increased both $\mt$ and $\er$, and that participants also subjectively perceived stronger blur as increasing difficulty.
We also derived several candidate models for predicting $\mt$ while taking blur level into account, and showed that Equation~\ref{eq:fittsAB} performed best in terms of both fit to measured $\mt$ data and predictive accuracy for untested conditions.
Using this model, we further showed in a follow-up experiment that correcting target size can partially reduce the effect of increasing blur level on $\mt$.

The conventional assumption, which has often remained implicit in prior work, that participants can clearly see both the targets and the cursor on the screen, does not necessarily hold in diverse real-world usage environments.
Showing that even under such visually difficult conditions, $\mt$ can still be estimated with high accuracy by appropriately introducing blur level $B$ into Fitts' law represents an important theoretical extension in HCI.
We hope that this work will contribute to future improvements in accessibility and to the development of UI design support tools.


\section*{Funding}
This work was not supported by any grant.

\section*{Disclosure statement}
The fourth author is employed by LY Corporation.
There is no other potential conflict of interest.

\section*{Author Contributions}
Ryuto Tomihari contributed to Conceptualization, Data curation, Formal analysis, Investigation, Methodology, Software, Validation, Visualization, Writing--original draft, Writing--review \& editing.
Taiki Kinoshita contributed to Conceptualization, Data curation, Formal analysis, Investigation, Methodology, Software, Validation, Visualization, Writing--original draft, Writing--review \& editing.
Yosuke Oba contributed to Conceptualization, Data curation, Formal analysis, Investigation, Methodology, Software, Validation, Visualization, Writing--original draft, Writing--review \& editing.
Shota Yamanaka contributed to Conceptualization, Formal analysis, Methodology, Software, Validation, Visualization, Writing--original draft, Writing--review \& editing.
Homei Miyashita contributed to Conceptualization, Methodology, Writing--review \& editing.

\section*{Use of Large Language Models}
Generative AI (ChatGPT 5.4 Pro) was used to improve the quality of writing, including style, phrasing, and grammar, as well as to check the mathematical correctness of model derivations.
The study design, data collection, analysis, interpretation, and final responsibility for the manuscript remain with the authors.

\bibliography{zinteractapasample}

\appendix

\section*{Biography}

\noindent\textbf{Ryuto Tomihari} received his master's degree from Meiji University in 2024. His research interests include human-computer interaction, pointing performance, accessibility, and human performance modeling.

\noindent\textbf{Taiki Kinoshita} received his master's degree from Meiji University in 2024. His research interests include human-computer interaction, pointing techniques, and experimental methodology.

\noindent\textbf{Yosuke Oba} received his master's degree from Meiji University in 2024. His research interests include human-computer interaction, pointing performance on graphical user interfaces, and performance modeling.

\noindent\textbf{Shota Yamanaka} is a senior chief researcher at LY Corporation Research, LY Corporation. He received his Ph.D. in engineering from Meiji University in 2016. His research interests include human-computer interaction, graphical user interfaces, and human performance modeling.

\noindent\textbf{Homei Miyashita} is a professor and chair of the department of Frontier Media Science, School of Interdisciplinary Mathematical Sciences, Meiji University. He specializes in human-computer interaction and has recently developed taste media that can record and reproduce taste.

\end{document}